\documentclass[journal,comsoc]{IEEEtran}
\usepackage{cite}
\usepackage{microtype}
\usepackage{amsmath}
\usepackage{accents}
\usepackage{amssymb}
\usepackage{graphicx}
\usepackage{suffix}
\usepackage{mathtools}
\usepackage{stackengine}
\usepackage{caption}
\usepackage{subcaption}
\usepackage[export]{adjustbox}
\def\delequal{\mathrel{\ensurestackMath{\stackon[1pt]{=}{\scriptstyle\Delta}}}}

\usepackage[square,sort,comma,numbers]{natbib}

\newcommand{\meijerG}[7]{\text G^{#1,#2}_{#3,#4} \left. \Big[ #7 \middle\vert \right.  \begin{smallmatrix} #5 \\ #6 \end{smallmatrix} \Big]}

\newcommand{\addtag}{\refstepcounter{equation}\tag{\theequation}}

\usepackage{scalerel,stackengine}

\newcommand\underrel[2]{\mathrel{\mathop{#2}_{#1}}}
\begin{document}
\title{{\huge Interference-Limited Mixed MUD-RF/FSO Two-Way Cooperative Networks over Double Generalized Gamma Turbulence Channels}}
\author{Abhijeet Upadhya, \and Vivek K. Dwivedi, \textit{Member, IEEE,} \and  and Mohamed-Slim Alouini,\textit{ Fellow, IEEE}  \thanks{Abhijeet Upadhya and Vivek K. Dwivedi are with department of Electronics and Communication Engineering, Jaypee Institute of Information Technology, Noida, India, (Email: upadhya.abhijeet@gmail.com, vivek.dwivedi@jiit.ac.in).} \thanks{Mohamed-Slim Alouini is with Computer, Electrical, and Mathematical Sciences and Engineering Division at King Abdullah University of Science and Technology (KAUST), Thuwal, Makkah Province, Saudi Arabia (email: slim.alouini@kaust.edu.sa).}	
}
\maketitle
	\begin{abstract}
		In this letter, the performance of multiuser-radio frequency/free space optics (RF/FSO) two-way relay network in the presence of interference is investigated. The FSO link accounts for pointing errors and both types of detection techniques, i.e. intensity modulation/direct detection as well as coherent demodulation, which is modeled as double generalized gamma (D-GG) turbulence channel. On the other hand, the multiple users on the RF link are assumed to undergo Nakagami-\(m\) fading. Multiple co-channel interferers (CCIs) which corrupt the signal at relay node are modeled using Nakagami-\(m\) distribution. Specifically, the exact closed-form expressions for the outage probability (OP) of the overall system is derived. Moreover, the closed form expression for the achievable sum-rate (ASR) of the considered system is presented. In order to simplify the results, the asymptotic approximations of the OP and ASR are derived in terms of elementary functions. The results presented in the paper are validated by Monte-Carlo simulations.
	\end{abstract}
\begin{IEEEkeywords}
	Free-space optical (FSO) communication, multiuser diversity, co-channel interference, Fox's H-functions.
\end{IEEEkeywords}

\section{Introduction}
The offerings of FSO systems such as rapid deployment time, high security, flexibility are limited by the atmospheric turbulence and pointing error \cite{E.Lee}. Besides, utilization of FSO system as an alternative and/or a complement to radio frequency (RF) counterparts has given rise to mixed RF/FSO relay systems \cite{E.Lee, DGGwithPointError}. Relaying schemes provide benefits such as spatial diversity gains without increasing the receiver hardware, uniform quality-of-service and extended coverage. Conventional one-way-relaying (OWR) is spectrally inefficient since the exchange of information between source to destination requires two complete time-slots. Spectral efficiency can be improved by utilizing two-way-relaying (TWR) scheme \cite{TWR-RF}, where the transmission of information from two nodes takes place simultaneously, through a relay node. In the first phase, the two nodes transmit information simultaneously to the relay node, which is broadcasted by the relay to the designated destinations in the second phase of communication. This brings about an improvement by factor of 2 when compared to OWR strategy. \par 
While most of the known results for the FSO relay-assisted communications have relied on the absence of interference, it is important to note that asymmetric mixed RF/FSO relay systems are inherently vulnerable to the effect of co-channel interference (CCI) due to the involvement of RF links. The recognition of the interference-limited behavior of mixed RF/FSO systems has motivated the authors of \cite{RF-FSO-Rel-Det_Intf} to derive the expression of outage probability (OP) and bit error rate (BER) for an interference limited mixed RF/FSO amplify-and-forward (AF) OWR relaying system. The interfering signals have been modeled using the Nakagami-\(m\) fading distribution, while the FSO systems are assumed to follow double-generalized Gamma (D-GG) turbulence model to demonstrate exact and asymptotic performance of the overall system in \cite{RF-FSO-Rel-Det_Intf}. Moreover, the effect of multiple CCIs on the mixed RF/FSO OWR systems have been analyzed in \cite{RF-FSO-CSI-AF} where the authors have derived expressions of the OP, BER and capacity for channel state information (CSI) assisted system. On the other hand, impact of multiuser diversity (MUD) on interference limited mixed RF/FSO OWR systems has been indicated by the authors of \cite{mypaper2}. Furthermore, research work \cite{TWR-MUD} demonstrates the improvement imparted by MUD scheme on mixed RF/FSO TWR system for interference free transmission. Notably, no work has been reported on the effect of interference over MUD assisted mixed RF/FSO TWR system. This motivates to explore the performance for a bidirectional MUD-RF/FSO system, where the relay node operates in the presence of multiple CCIs. The main contributions of this work are: 
\begin{enumerate}
	\item In this work, performance analysis of interference limited RF/FSO cooperative TWR networks is performed, where user diversity scheme is implemented, such that the relay node selects the user with best channel condition.
	\item The closed form expressions for performance metrics such as OP and ergodic sum-rate is derived.
	\item The OP and ergodic sum-rate expressions account for both intensity modulation/direct detection (IM/DD) and coherent demodulation schemes on the optical link, in the presence of pointing error.
	\item Finally, the derived closed form expression for the OP is expressed asymptotically to validate the proposed work.
\end{enumerate}
\section{SYSTEM AND CHANNEL MODELS}
Consider a  mixed RF/FSO two-way cooperative AF relaying system, where \(K\) mobile users on the RF link communicate with the FSO destination node through the bidirectional relay node \(R\). In particular, the relay node exchanges information between two source nodes \(S_{\scaleto{RF,j}{5 pt}}\) and \(S_{\scaleto{FSO}{5 pt}}\), where \(S_{\scaleto{RF,j}{5 pt}}\) represents the \(j^{th}\)-user selected by the relay node based on best channel conditions. The user selection at relay node is performed using the opportunistic scheduling based on the quality of the \(S_{\scaleto{RF,j}{5 pt}} \rightleftarrows R \) links. It is assumed that \(N\) number of interferers following Nakagami-\(m\) distribution, corrupt the signal at relay node. In the multiple access phase \(T_1\) of communication, both RF and FSO nodes simultaneously transmit their respective information to relay node \(R\). The received  signal \(y_{\scaleto{R}{5 pt}}\) at node \(R\) is given by
\begin{align*} %(Eq. 1)
y_{R} = \sqrt{P_{\scaleto{RF,j}{3 pt}}} h_{\scaleto{RF, j}{3 pt}} x_{\scaleto{RF}{3 pt}} + \sqrt{P_{\scaleto{FSO}{3 pt}}^{\scaleto{Opt}{3 pt}}} g_{\scaleto{FSO}{3 pt}} x_{\scaleto{FSO}{3 pt}}+ \sum_{i=1}^{N} \sqrt{P_{\scaleto{I_{R,i}}{3 pt}}}  h_{\scaleto{I,r}{3 pt}} x_{r,i}  + N_{\scaleto{T_1}{4 pt}} \addtag \label{sysEq1}
\end{align*}
where \(h_{\scaleto{RF, j}{5 pt}}\) is Nakagami-\(m\) distributed fading amplitude of RF channel carrying information \(x_{\scaleto{RF}{5 pt}}\) of \(j^{th}\)-user and \(g_{\scaleto{FSO}{4 pt}} \) represents irradiance fluctuation on the FSO link. The \(P_{\scaleto{RF,j}{5 pt}}\) stands for useful transmit power whereas \(P_{\scaleto{I_R}{5 pt}}\) denotes the interference power with channel coefficient \(h_{I_r} \) at the relay node \(R\), \(x_{r,i}\) is the symbol emitted by \(i^{th}\) interferer, and \(N_{T_1}\) represents the additive white Gaussian noise (AWGN).  
In the second phase of communication, the relay node \(R\) sends the processed version of the signal \(x_{\scaleto{RF}{5 pt}}\) towards the FSO node \(S_{\scaleto{FSO}{5 pt}}\) and symbol \(x_{\scaleto{FSO}{5 pt}}\) to \(S_{\scaleto{RF,j}{5 pt}}\). Specifically, the relay node amplifies the received signal by \(G = \frac{1}{|h_{\scaleto{RF, j}{5 pt}}|^2+|g_{\scaleto{FSO}{4 pt}}|^2} \), and forwards it to \(S_{\scaleto{FSO}{5 pt}}\) and \(S_{\scaleto{RF,j}{5 pt}}\). It is considered that, the channels \(S_{\scaleto{RF,j}{5 pt}} \rightleftarrows R \) and \(S_{\scaleto{FSO}{5 pt}} \rightleftarrows R \) are reciprocal and that the channel coefficients remain unchanged during both the phases of transmission. Given that the two nodes can perform self-interference cancellation with the knowledge of  \(h_{\scaleto{RF, j}{5 pt}} \) and \(g_{\scaleto{FSO}{5 pt}}\), the received signal at RF node can be estimated as:
\begin{align*} %(Eq. 2)
\tilde{y}_{\scaleto{T_2, RF}{4 pt}}  = &  \sqrt{P_{\scaleto{R}{3 pt}}}(\eta g_{\scaleto{FSO}{2 pt}})^{\frac{r}{2}} G  h_{\scaleto{RF, j}{3 pt}} x_{\scaleto{FSO}{3 pt}}  + \sqrt{P_{\scaleto{R}{3 pt}}}  h_{\scaleto{RF, j}{3 pt}}  G \sum_{i=1}^{N} \sqrt{P_{\scaleto{I_{R,i}}{3 pt}}}  h_{\scaleto{I_r}{3 pt}} x_{\scaleto{r,i}{3 pt}} + N_{\scaleto{01}{3 pt}} 
\end{align*}
where \(P_{R}\) is the power of signal transmitted from relay node, the overall noise from \(S_{\scaleto{FSO}{5 pt}}\) and \(S_{\scaleto{RF,j}{5 pt}}\) is represented as \(N_{01}\), whereas \(\eta\) is the electrical-to-optical conversion coefficient. Constant \(r\) denotes the type of optical demodulation employed where \(r=1\) represents coherent demodulation and \(r=2\) corresponds to IM/DD demodulation. Similarly, the estimated signal at the FSO node can be given as:
\begin{align*} %(Eq. 3)
\tilde{y}_{\scaleto{T_2, FSO}{4 pt}}  =  & \sqrt{P_{\scaleto{R}{3 pt}}}(\eta g_{\scaleto{FSO}{3 pt}})^{\frac{r}{2}}  G  h_{\scaleto{RF, j}{3 pt}} x_{RF}  + g_{\scaleto{FSO}{3 pt}} \sqrt{P_{\scaleto{R}{3 pt}}} G \sum_{i=1}^{N} \sqrt{P_{\scaleto{I_{R,i}}{3 pt}}}  h_{\scaleto{I_r}{3 pt}} x_{\scaleto{r,i}{3 pt}} + N_{\scaleto{02}{3 pt}} 
\end{align*}
where \(N_{02}\) represents the overall noise from \(S_{\scaleto{RF,j}{5 pt}}\) and \(S_{\scaleto{FSO}{5 pt}}\) node. Without loss of generality, the noise contributions \(N_{01}\) and \(N_{02}\) are assumed to be AWGN with zero-mean and equal variances \({\sigma_n}^2\). The end-to-end signal-to-interference-plus-noise ratio (SINR) for \(S_{\scaleto{RF,j}{5 pt}} \rightarrow R \rightarrow  S_{\scaleto{FSO}{5 pt}} \) can be given as
\begin{align*} %(Eq. 4)
\gamma_{\scaleto{{T_2,RF}}{4 pt}} = \frac{P_R G^2 |h_{\scaleto{RF, j}{5 pt}}|^2 |g_{\scaleto{FSO}{3 pt}}|^2}{\sum_{i=0}^{N} {P_{I_{R,i}}}  |h_{I_r}|^2  |g_{\scaleto{FSO}{3 pt}}|^2 + {\sigma_n}^2} \addtag \label{(T2, RF)}
\end{align*}
 Similarly, the end-to-end SINR for \(S_{\scaleto{FSO}{5 pt}}\rightarrow R\rightarrow  S_{\scaleto{RF,j}{5 pt}}\) can be formulated as 
\begin{align*} %(Eq. 5)
\gamma_{\scaleto{{T_2,FSO}}{4 pt}} = \frac{P_R G^2 |h_{\scaleto{RF, j}{5 pt}}|^2 |g_{\scaleto{FSO}{3 pt}}|^2}{\sum_{i=0}^{N} {P_{I_{R,i}}}  |h_{I_r}|^2  |h_{\scaleto{RF, j}{5 pt}}|^2 + {\sigma_n}^2 } \addtag \label{(T2, FSO)}
\end{align*}
\par
Under the assumption of Nakagami-\(m\) distribution, the fading coefficients on the RF link, for the \(k^{th}\) user, follows the probability density function (PDF) given by \cite{wireleDigAloiuni} \(f_{\gamma_k} (\gamma) = \frac{\beta^{m_{RF}}}{\Gamma(m_{RF})} \gamma^{m_{RF}-1}\exp(-\beta\gamma) \), where \(m_{RF}\) is the shape parameter, \(\beta=\frac{m_{RF}}{\bar\gamma_{k}} \) and \(\bar\gamma_{k}\) is the average SNR of the \(k^{th}-\)user. Furthermore, the PDF of instantaneous SNR on the RF link with \(K\) users can be obtained using the definition of ordered statistics as \(f_{\gamma_{SR},j}(\gamma) = K [F_{\gamma_{k}(\gamma)}]^{K-1}f_{\gamma_k} (\gamma)\). The closed form expression for the cumulative distribution function (CDF) can be obtained using \(F_{\gamma_{\scaleto{{RF}}{4 pt}}}(\gamma) = \int_{0}^{\gamma} f_{\gamma_{SR},j}(y) dy \). For integer values of \(m_{RF}\), utilizing \cite[Eq. (2)]{mypaper2} along with \cite[Eq. (3.381.1) and (8.352.1)]{RyzhikTables}, with some mathematical manipulations, the closed-form expression for the CDF can be expressed as:
\begin{align*}%(Eq. 12)
&F_{\gamma_{\scaleto{{RF}}{3 pt}}}(\gamma) =  \sum_{n_1=0}^{K-1} \sum_{n_2=0}^{n_1(m_{\scaleto{RF}{2 pt}}-1)} A_1 \Bigg(1- e^{(-\mathcal{B}_0\gamma)}\sum_{l=0}^{m_{\scaleto{RF}{2 pt}}+n_2-1} \frac{{(\mathcal{B}_0\gamma)}^{l}}{l!} \Bigg)  \addtag \label{RF-Eff-CDF}
\end{align*}
where \(\mathcal{B}_0 = \beta(n_1+1) \) and \(A_1\) is given by: %(Eq. 13)
\[ A_1 = \frac{K(m_{\scaleto{RF}{2 pt}}+n_2-1)!}{\Gamma m_{\scaleto{RF}{4 pt}}} (-1)^{n_1} \binom{K-1}{n_1} \zeta_{n_{1}n_{2}}(m_{\scaleto{RF}{4 pt}}) (n_1+1)^{-n_2-m_{\scaleto{RF}{4 pt}}} \addtag  \]
 where \(\zeta_{n_{1}n_{2}}(m_{\scaleto{RF}{3 pt}})\) is the coefficient of multinomial expansion which can be recursively obtained using the relation \(\zeta_{n_{1}n_{2}}=\sum_{b=n_1 -x+1}^{n_1}\frac{\zeta_{bn_{2}-1}}{n_1-b} \text I_{\scaleto{[0,(n_2-1)(x-1)]}{4 pt}} \) where \(I_{\scaleto{[a,b]}{4 pt}}\) is the indicator function as defined in \cite[Eq. (9.120)]{wireleDigAloiuni}. 
 \par
\begin{figure*}[!t] % Eq (14) Outage Probability }{
	\begin{align*}
	\setcounter{equation}{12}
	P_{\scaleto{out}{4 pt}} & =   \sum_{n_1=0}^{K-1} \sum_{n_2=0}^{n_1(m_{RF}-1)} A_2 \Bigg[\frac{y^{Nm_{1}}}{(2\pi)^{\frac{y-1}{2}}} \meijerG{n}{y+1}{y+1+r}{n+1}{1, \tau_3, \tau_5}{\tau_4, 0}{ \mathcal{D}_7 \gamma_{{\scaleto{th}{3 pt}}}^y} -  \sum_{m=0}^{n_2+m_{{\scaleto{RF}{4 pt}}}-1} \frac{(\mathcal{B}_0 \gamma_{{\scaleto{th}{3 pt}}})^m}{m!} \left(\frac{\Omega_I{_1}}{m_1}\right)^{-m}  
	\text{H}_{\mathbf{x}_1}^{\mathbf{x}_2}\left[
	\begin{array}{c}
	\tau_{6}  \\
	\noindent\rule{0.4cm}{0.5pt}
	\end{array}\middle\vert
	\begin{array}{c}
	\noindent\rule{0.4cm}{0.5pt}\\
	(0,1)\\
	\end{array}\middle\vert
	\begin{array}{c}
	\tau_{7}\\
	\tau_{8}\\
	\end{array}\middle\vert
	\mathcal{B}_1,  \mathcal{B}_2 \right] 
	\Bigg]
	\addtag \label{Outage}
	\end{align*}
	\rule{\linewidth}{1pt}
\end{figure*} 
On the other hand, it is assumed that \(R \rightleftarrows  S_{\scaleto{FSO}{5 pt}} \) link encompasses the turbulence-induced fading \(I_a\) with pointing errors \(I_p\) such that \(I=I_aI_p\).
The atmospheric turbulence fading \(I_a\) on the FSO link is modeled as double generalized gamma (D-GG) fading model \cite{DGGwithPointError}. The irradiance \(I_a=I_xI_y\), such that \(I_x\sim GG(\alpha_1, \beta_1, \Omega_1) \) and \(I_y\sim GG(\alpha_2, \beta_2, \Omega_2) \). \(\beta_1\) and \(\beta_2\) are shaping parameters, whereas \(\alpha_1\) , \(\Omega_1\), \(\alpha_2\) and \(\Omega_2\) are calculated based on the variances of the small and large scale fluctuations. The PDF of SNR on the FSO link can be expressed as \cite{DGGwithPointError}:
\begin{align*}%(Eq. 16)
\setcounter{equation}{7}
f_{\gamma_{\scaleto{FSO}{4 pt}}}(\gamma) = \frac{\mathcal{D}_1}{r \gamma} \meijerG{\lambda+\sigma+1}{0}{1}{\lambda+\sigma+1}{\tau_2}{\tau_1}{\mathcal{D}_2 z^{y} \left(\frac{\gamma}{\mu_r} \right)^{\frac{y}{r}} } \addtag \label{FSO-PDF}
\end{align*}
with \(\mathcal{D}_1 = \frac{\xi^2 \sigma^{\beta_1-\frac{1}{2}} \sigma^{\beta_2-\frac{1}{2}} (2\pi)^{1-\frac{\sigma+\lambda}{2}} }{\Gamma(\beta_1) \Gamma(\beta_2)} \), 
\( \mathcal{D}_2 = \frac{\beta_1^\sigma \beta_2^\lambda}{\lambda^\lambda\sigma^\sigma\Omega_1^{\sigma} \Omega_2^{\lambda}} \)
 where \(\tau_1 = \left[ \frac{\xi^2}{y}, \Delta(\sigma:\beta_1), \Delta(\lambda:\beta_2) \right] \), \(\tau_2=\left[1+\frac{\xi^2}{y}\right] \), \(\mu_r= \frac{\left(\eta \mathbb{E} (I)\right)^r}{N_0} \) and \(r\) denotes the kind of demodulation scheme \cite{DGGwithPointError}. Considering \(\mathcal{D}_3= \prod_{g=1}^{\sigma+\lambda} \Gamma\left(\frac{1}{y} + \tau_{\scaleto{0,g}{3.5 pt}}  \right) \), \(z\) can be given as \(z=\frac{\mathcal{D}_1(\mathcal{D}_2)^{1/y}\mathcal{D}_3}{(1+\xi^2)}  \). \\
 where \(\tau_0 = \left[\Delta(\sigma:\beta_1), \Delta(\lambda:\beta_2)\right] \) with \(\Delta(z:x)\) defined as \([\frac{x}{z}, \frac{x+1}{z}, \dots, \frac{x+z-1}{z}]\). Moreover, \(\meijerG{m}{n}{p}{q}{}{}{.}\) is the Meijer-G function defined in \cite[Eq. (9.301)]{RyzhikTables} and \(y=\alpha_2\lambda \). The pointing error parameter \(\xi \) is defined as the ratio between the equivalent beam width \(\omega_{eq}\) and pointing error jitter standard deviation \(\sigma_{s} \), given by relation \(\xi=\frac{\omega^{2}_{eq}}{\sigma_{s}} \) \cite{DGGwithPointError}.   \par
Further, the CDF of FSO channel over D-GG atmospheric turbulence with pointing errors can be formulated as follows \cite{DGGwithPointError}
\begin{align*}%(Eq. 10)
 F_{\gamma_{\scaleto{FSO}{4 pt}}}(\gamma)=\mathcal{D}_4  \meijerG{n}{1}{r+1}{n+1}{1,\tau_3}{\tau_4,0}{\mathcal{D}_5 \left(\frac{\gamma}{\mu_r} \right)^{y} } \addtag \label{FSO-CDF}
\end{align*}
where \(n=r(\lambda+\sigma+1) \), \(\mathcal{D}_4= \frac{\xi^2 \sigma^{\beta_1-\frac{1}{2}} \sigma^{\beta_2-\frac{1}{2}} (2\pi)^{1-\frac{r(\sigma+\lambda)}{2}} r^{(\beta_1+\beta_2-2)} }{y\Gamma(\beta_1) \Gamma(\beta_2)} \), \(\mathcal{D}_5= \left(\frac{\mathcal{D}_2 z^y }{r^{\sigma+\lambda}} \right)^r \), \(\tau_3= \left[\Delta(r: \tau_2) \right] \) and \(\tau_4= \left[\Delta(r: \tau_1) \right] \) comprising of \(r(\lambda+\sigma+1) \) terms. 
\section{PERFORMANCE ANALYSIS}
In this section, the investigation of outage performance and sum-rate of the bidirectional relay system based on the aforementioned channel models is presented. 
\subsection{Exact Outage Probability}
The outage probability (OP) is an important performance indicator of wireless communication system. OP is defined as the probability that the instantaneous SNR falls below a pre-defined threshold \(\gamma_{\scaleto{{th}}{4 pt}}\). For the considered TWR system, the OP can be defined as:
\begin{align*}
\setcounter{equation}{9}
P_{\scaleto{out}{4 pt}} = & F_{e_{2} e} (\gamma)|_{\gamma_{\scaleto{{th}}{3 pt}}}  = \text{Pr}\left(\text{min}[\gamma_{\scaleto{{T_2,RF}}{4 pt}}, \gamma_{\scaleto{{T_2,FSO}}{4 pt}}<\gamma_{\scaleto{{th}}{4 pt}}]\right) \\ & 
= 1- \text{Pr}\left(\text{min}[\gamma_{\scaleto{{T_2,RF}}{4 pt}}>\gamma_{\scaleto{{th}}{4 pt}}, \gamma_{\scaleto{{T_2,FSO}}{4 pt}}>\gamma_{\scaleto{{th}}{4 pt}}]\right)\\ & 
=1-\text{Pr}\left(\frac{P_{\scaleto{R}{4 pt}} \text{min}\left[|h_{\scaleto{RF, j}{5 pt}}|^2 , |g_{\scaleto{FSO}{3 pt}}|^2 \right] }{\sum_{i=0}^{N} {P_{I_{R,i}}}  |h_{I_r}|^2}  > \gamma_{\scaleto{{th}}{4 pt}} \right) 
\addtag \label{e2eCDF}
\end{align*}
 Moreover, the PDF of the total interference-to-noise ratio (INR) \(\sum_{i=1}^{L} \gamma_{\scaleto{I_{r}}{4 pt},i}\) can be expressed as \cite{wireleDigAloiuni}:
\begin{align*} %(Eq. 9) Intf-RF
f_{_{{\scaleto{I_{r}}{5 pt}}}}(\gamma)& = \bigg[\frac{m_{1}}{\Omega_I{_1}}\bigg]^{m_{1}L} \frac{\gamma^{m_{1}L-1}}{\Gamma(m_{1}L)}\exp\bigg(-\frac{m_{1}L}{\Omega_I{_1}}\gamma\bigg) \addtag \label{Intf-R}
\end{align*}
where \(m_1\) is Nakagami-\(m\) fading parameter and \(\Omega_I{_1}\) is the average interference to noise ratio (INR) at the relay node. Defining \(Y \delequal P_{\scaleto{R}{4 pt}} \text{min}\left[|h_{\scaleto{RF, j}{5 pt}}|^2 , |g_{\scaleto{FSO}{3 pt}}|^2 \right] \), and considering statistical independence between \(\gamma_{\scaleto{{T_2,RF}}{4 pt}}\) and \(\gamma_{\scaleto{{T_2,FSO}}{4 pt}}\), the OP can be further calculated as:
\begin{align*}
P_{\scaleto{out}{4 pt}} & = 1-  \int_{0}^{\infty} \text{Pr}\left( Y > z\gamma_{\scaleto{{th}}{4 pt}}  \right) f_{_{{\scaleto{I_{r}}{5 pt}}}}(z) dz \\ & =
1-  \int_{0}^{\infty} F_{\gamma_{\scaleto{{RF}}{4 pt}}}\left(  z\gamma_{\scaleto{{th}}{4 pt}}  \right) F_{\gamma_{\scaleto{{FSO}}{4 pt}}}\left(  z\gamma_{\scaleto{{th}}{4 pt}}  \right) f_{_{{\scaleto{I_{r}}{5 pt}}}}(z) dz \addtag \label{Pout-Int}
\end{align*}
Appropriately substituting (\ref{RF-Eff-CDF}), (\ref{FSO-CDF}) and (\ref{Intf-R}) into (\ref{Pout-Int}), while invoking \cite[Eq. (2.3)]{FoxHIntgrl}, with some mathematical manipulations, closed-form expression for the OP is derived as given in (\ref{Outage}), where \(\text{H}_{\mathbf{h}_1}^{\mathbf{h}_2}\left[
\mathcal{A}_1,  \mathcal{A}_2 \right] \) is the bivariate Fox's H-function as defined in \cite[Eq. (1.1)]{FoxHIntgrl}. The order of H-function can be formulated as \(\{\mathbf{x}_1, \mathbf{x}_2\} = \{(\text 0, \text 1:\text 1,\text 0:\text {n}, \text {1}) , (\text 1, \text 0:\text 0,\text 1:\text {r+1}, \text {n+1})\} \). Moreover, in (\ref{Outage}), the argument of Meijer-G is given by \(\mathcal{D}_7=\mathcal{D}_5 \left(\frac{y\Omega_I{_1}}{m_1\mu_{\scaleto{r}{3 pt}}} \right)^{y}\), whereas the arguments of Fox's H-function are derived to be \(\mathcal{B}_1 = \frac{\Omega_I{_1}}{m_1}\mathcal{B}_0 \gamma_{{\scaleto{th}{3 pt}}} \) and \(\mathcal{B}_2 = \mathcal{D}_5 \left(\frac{\Omega_I{_1} \gamma_{{\scaleto{th}{3 pt}}}}{\mu_{\scaleto{r}{3 pt}}m_1}\right)^y \). The constant \(A_2 = A_1 \mathcal{D}_4 \frac{(m_{\scaleto{RF}{3 pt}}+n_2-1)!}{\Gamma(m_1N)} \), while various parameters involved in (\ref{Outage}) can be defined as \(\tau_5 = \left[\Delta(y:1-m_1N)\right]\), \(\tau_6 = \left(1-m-m_1N, 1: y \right)\), \(\tau_7 =\left[ (1, 1), (\tau_3, [1]_{\text{length}(\tau_3)} )  \right] \) and \(\tau_8 =\left[(\tau_4, [1]_{\text{length}(\tau_4)} ), (0, 1) \right] \). The bivariate Fox's-H function can be evaluated numerically using the efficient MATLAB implementation as provided in \cite{BivFoxMATLAB}.
%%%%%%%%%%%%%%%%%%%%%%%%%%%%%%%%%%%%%%%%%%%%%%%%%%%%%%%%%%%%%%%%%%%%%%%%%%%%%%%%%%%%%%%%%
\iffalse % This is another way of presenting three figures on the one colomn frmat
%%%%%%%%%%%%%%%%%%%%%%%%%%%%%%%%%%%%%%%%%%%%%%%%%%%%%%%%%%%%%%%%%%%%%%%%%%%%%%%%%%%%%%%%% 
\begin{figure*}
	\begin{subfigure}[t]{0.45\textwidth}
		\includegraphics[width=\linewidth]{out1.eps}
		\caption{Effect of different atmospheric turbulence conditions and strength of interfering signals on outage probability performance of dual hop mixed RF/FSO TWR relaying system with interference on heterodyne detection.}
		\label{fig:out11}
	\end{subfigure}%
	\begin{subfigure}[t]{0.45\textwidth}
		\includegraphics[width=\linewidth]{out2.eps}
		\caption{Effect of type of demodulation, number of interferers, \(N\), and pointing error on outage probability performance of dual hop mixed RF/FSO TWR relaying system with interference.}
		\label{fig:out22}
	\end{subfigure}
\end{figure*}
%%%%%%%%%%%%%%%%%%%%%%%%%%%%%%%%%%%%%%%%%%%%%%%%%%%%%%%%%%%%%%%%%%%%%%%%%%%%%%%%%%%%%%%%% 
\fi

	\begin{figure}[!h]
	\centering
	\includegraphics[width=8.75cm,height=5.2cm]{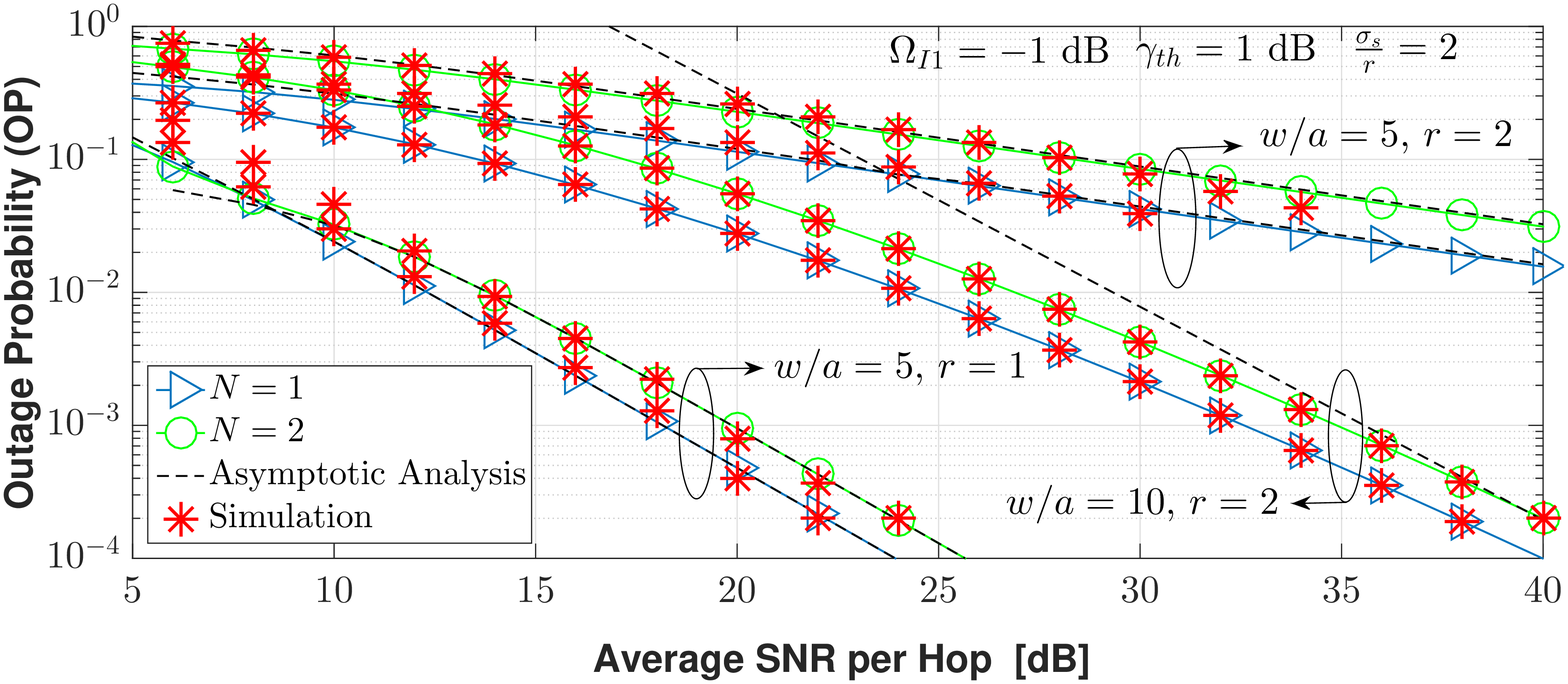}
	\caption{Effect of type of demodulation, number of interferers, \(N\), and pointing error on outage probability performance.}\label{fig:out1}
	\centering
	\includegraphics[width=8.5cm,height=5.2cm]{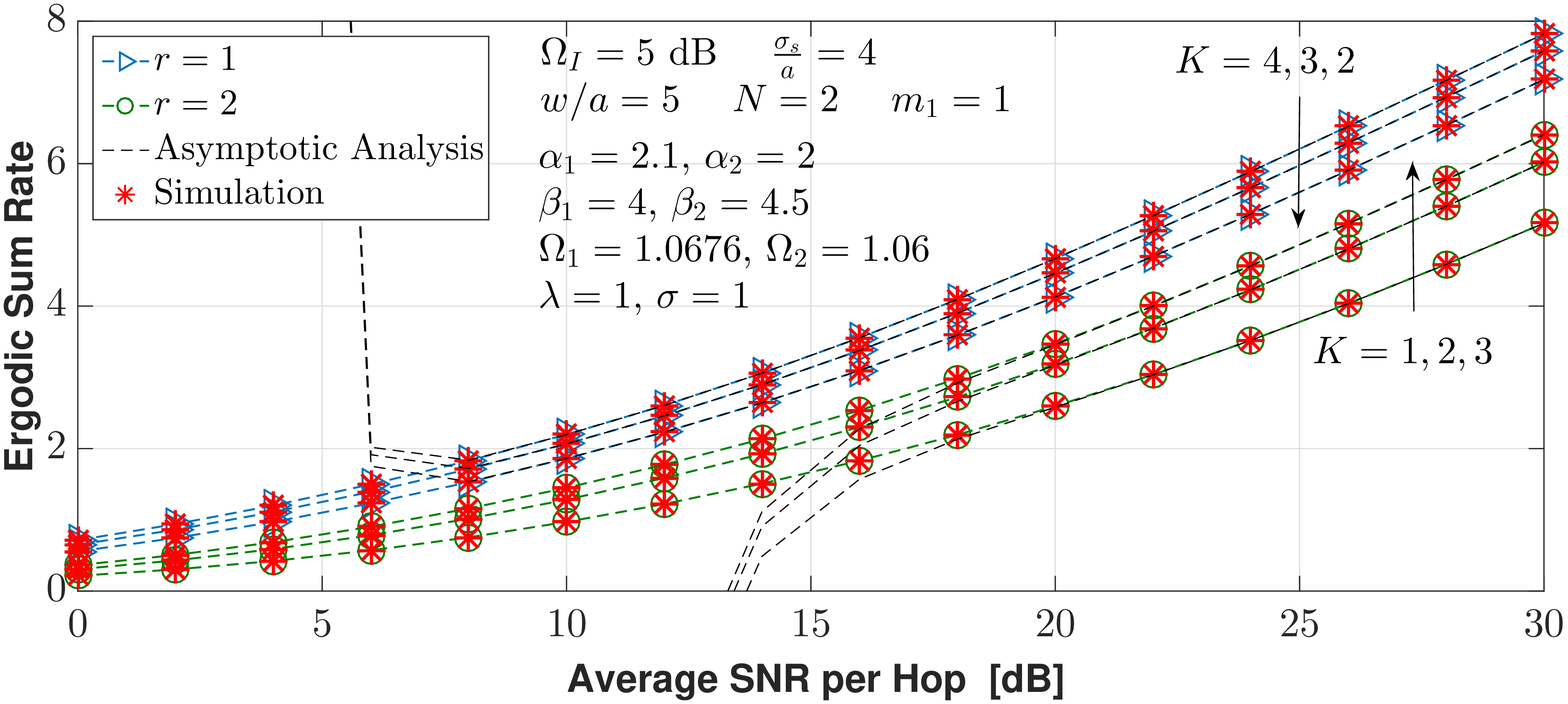}
	\caption{Effect of type of demodulation and number of users, \(K\), on the achievable ergodic sum rate.}\label{fig:cap1}
	\centering
\includegraphics[width=8.5cm,height=5.2cm]{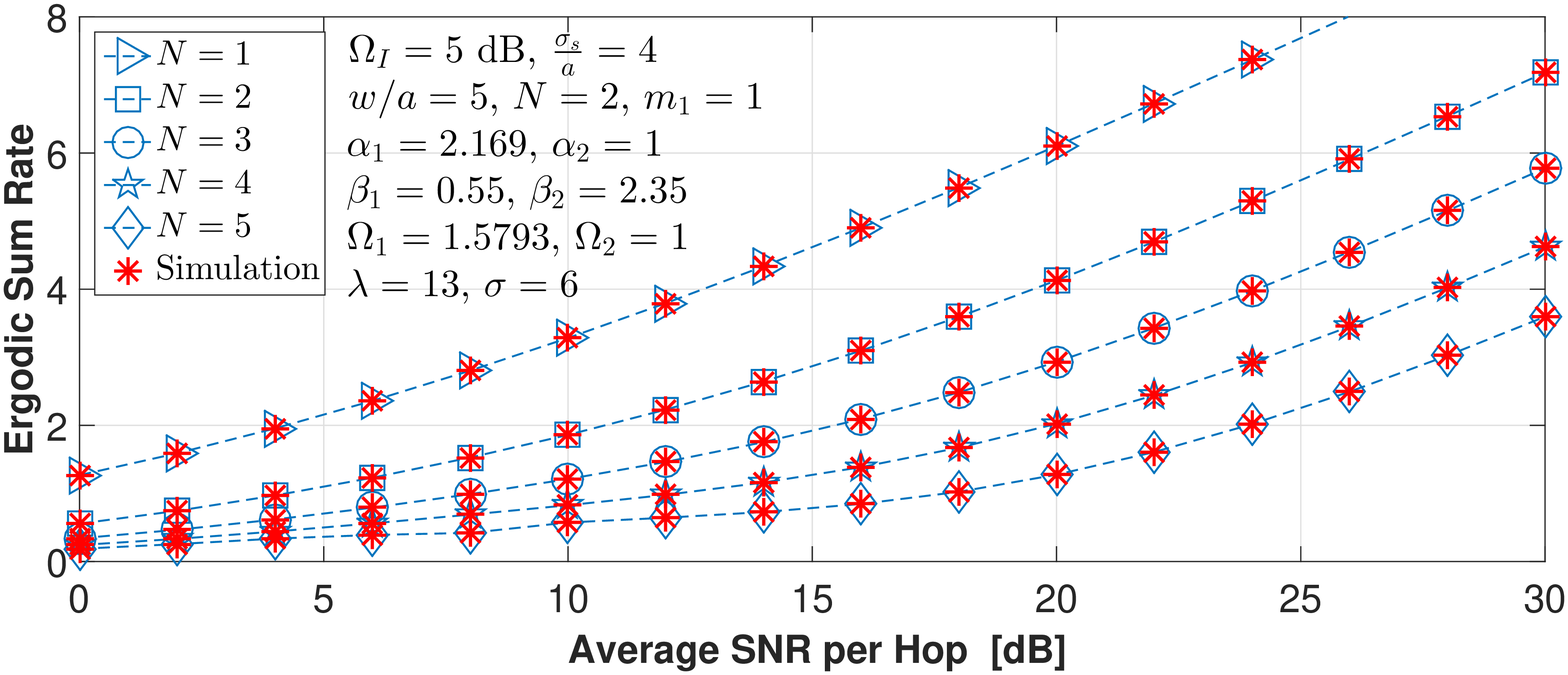}
\caption{Effect of varying number of interferers, \(N\), on the achievable ergodic sum rate of the two-way relaying networks.}\label{fig:cap2}
\end{figure}
\subsection{Asymptotic Outage Probability}
The exact expression derived in (\ref{Outage}) fails to offer quick insights into the performance of overall system. A simpler asymptotic (high SNR) expression for OP can be developed to get more insights about system's performance. According to \cite[Theorem (1.7) and Theorem (1.11)]{ResTheom}, the asymptotic expansion of H-function can be obtained as the residue of complex integration at the poles nearest to contour of integration. Assuming that \(\mathbf{p_n}=\min(\tau_4) \) is the dominant pole closest to contour, the expansion of Meijer-G function can be formulated as:
\begin{align*}
\setcounter{equation}{13}
& \meijerG{n}{y+1}{y+1+r}{n+1}{1, \tau_3, \tau_5}{\tau_4, 0}{ \mathcal{D}_7 \gamma_{{\scaleto{th}{3 pt}}}^y} \underrel{\mu_{\scaleto{r}{3 pt}} \to \infty}{\simeq}   \Lambda_1 {\prod_{j=1}^{y}{\Gamma(1-\tau_{5,j}-\mathbf{p_n})}} \gamma_{{\scaleto{th}{3 pt}}} ^{-y\mathbf{p_n}} \addtag \label{M-exp}
\end{align*}
where \(\Lambda_1= \frac{ {\mathcal{D}_{7}}^{-\mathbf{p_n}} \prod_{j=1}^{n}{\Gamma(\tau_{4,j}+\mathbf{p_n})}}{\mathbf{p_n} {\prod_{j=y+2}^{y+1+r}{\Gamma(\tau_{3,j}+\mathbf{p_n})}}}\). Similarly, by expressing the bivariate Fox's H-function in complex integral form using \cite[Eq. (1.1)]{FoxHIntgrl}, and making use of identity \cite[Eq. (9.113)]{RyzhikTables}, the asymptotic representation can be given as:
\begin{align*}
&\text{H}_{\mathbf{x}_1}^{\mathbf{x}_2}\left[
\begin{array}{c}
\tau_{6}  \\
\noindent\rule{0.4cm}{0.5pt}
\end{array}\middle\vert
\begin{array}{c}
\noindent\rule{0.4cm}{0.5pt}\\
(0,1)\\
\end{array}\middle\vert
\begin{array}{c}
\tau_{7}\\
\tau_{8}\\
\end{array}\middle\vert
\mathcal{B}_1,  \mathcal{B}_2 \right] \underrel{\mu_{\scaleto{r}{3 pt}} \to \infty}{\simeq} \Gamma\left(m+m_1N-y\mathbf{p_n} \right) \\ &
\times {}_1F_0 \Big(m+m_1N-y\mathbf{p_n}; ; -\frac{\Omega_I{_1}\mathcal{B}_0}{m_1}\gamma_{{\scaleto{th}{3 pt}}} \Big)  \Lambda_1 \left\{\frac{\gamma_{{\scaleto{th}{3 pt}}}}{y} \right\} ^{-y\mathbf{p_n}} \addtag \label{H-func-Asym}
\end{align*}
where \({}_1F_0 \Big(a; ; b \Big) \) is the Gaussian hypergeometric function as defined in \cite[Eq. (9.111)]{RyzhikTables}. Plugging (\ref{M-exp}) and (\ref{H-func-Asym}) into (\ref{Outage}), the asymptotic high SNR approximation of OP can be obtained. 
\subsection{Achievable Sum Rate}
In this section, the achievable sum-rate (ASR) offered by the proposed model is derived. The sum-rate of the system can be defined as the throughput over all the channel realizations. The ASR of wireless fading channels can be defined as:
\[{\mathcal{R}}= \frac{1}{2} \mathbb{E}(1+\gamma_{\scaleto{{T_2,RF}}{4 pt}})+\frac{1}{2} \mathbb{E}(1+\gamma_{\scaleto{{T_2,FSO}}{4 pt}}) = {\mathcal{R}}_1+{\mathcal{R}}_2 \label{ASR_def} \addtag \]
where \(\mathbb{E}(.)\) denotes the expectation operator. The transmission rate for the link \(S_{\scaleto{RF,j}{5 pt}}\rightarrow R \rightarrow  S_{\scaleto{FSO}{5 pt}} \) can be obtained as:
\begin{align*}
{\mathcal{R}}_1 = \frac{1}{2} \int_{0}^{\infty} \text{log}_2(1+\gamma) f_{\gamma_{\scaleto{{T_2,RF}}{4 pt}}}(\gamma) d\gamma \addtag \label{R1-form}
\end{align*}
Firstly, a closed-form expression for \(F_{\gamma_{\scaleto{{T_2,RF}}{4 pt}}}(\gamma)\) can be attained using \( F_{\gamma_{\scaleto{{T_2,RF}}{4 pt}}}(\gamma) = \int_{0}^{\infty} F_{\gamma_{\scaleto{{RF}}{4 pt}}}(\gamma y) f_{_{{\scaleto{I_{r}}{5 pt}}}}(y) dy \). Further to this, placing (\ref{RF-Eff-CDF}) and (\ref{Intf-R}) and applying \cite[Eq. (3.381.4)]{RyzhikTables}, \(F_{\gamma_{\scaleto{{T_2,RF}}{4 pt}}}(\gamma)\) can be derived, which can be differentiated to obtain \(f_{\gamma_{\scaleto{{T_2,RF}}{4 pt}}}(\gamma) \), which can be represented as:
\begin{align*}
& f_{\gamma_{\scaleto{{T_2,RF}}{4 pt}}}(\gamma)  = \frac{1}{2} \sum_{n_1=0}^{K-1} \sum_{n_2=0}^{n_1(m_{\scaleto{RF}{3 pt}}-1)} \sum_{m=0}^{m_{\scaleto{RF}{3 pt}}+n_2-1} A_3 \Bigg[ \frac{(m+m_1N) {\mathcal{B}_0}}{m!} \gamma^{m} \\ & \times \Big[{\mathcal{B}_0}\gamma+\frac{m_1}{\Omega_I}\Big]^{-(m+m_1N+1)}
-m\gamma^{m} \Big[{\mathcal{B}_0}\gamma+\frac{m_1}{\Omega_I}\Big]^{-(m+m_1N)}  \Bigg] \addtag \label{PDF-Cap-1}
\end{align*}
where \(A_3 = A_1  \frac{(m_{\scaleto{RF}{3 pt}}+n_2-1)!}{\Gamma(m_1N)}\left(\frac{m_1}{\Omega_I} \right)^{m_1N} \frac{\Gamma(m+m_1N) {\mathcal{B}_0}^{m}}{m!} \). After representing \(\text{log}_2(1+\gamma) = \meijerG{1}{2}{2}{2}{1, 1}{1,1}{\gamma} \) with the aid of \cite[Eq. (07.34.03.0456.01)]{Wolfram}, substituting \(f_{\gamma_{\scaleto{{T_2,RF}}{4 pt}}}(\gamma)\) into (\ref{R1-form}) and applying \cite[Eq. (07.34.03.0271.01) and (07.34.21.0013.01)]{Wolfram}, with some mathematical manipulations, the closed-form expression of \(\mathcal{R}_1\) can be obtained as given in below: 
\begin{align*}
\setcounter{equation}{18}%%%%Expression of R1
{\mathcal{R}}_1 =   \frac{1}{2} \sum_{n_1=0}^{K-1} \sum_{n_2=0}^{n_1(m_{\scaleto{RF}{3 pt}}-1)} \sum_{m=0}^{m_{\scaleto{RF}{3 pt}}+n_2-1} A_3 & \Bigg[\left( \frac{ \mathcal{B}_0 \Omega_I{_1}}{m_1\delta}\right) \meijerG{3}{2}{3}{3}{\tau_{9}}{\tau_{10}}{ \mathcal{B}_{3} }\\ & -m\meijerG{3}{2}{3}{3}{\tau_{11}}{\tau_{12}}{ \mathcal{B}_{3} } \Bigg]  \label{R1-Cap} \addtag
\end{align*}
where \(\mathcal{B}_3= \frac{\Omega_1\mathcal{B}_0}{m_1 \delta} \), whereas the parameters are defined as:
\(\tau_9 =\left[-m-m_1N, -m-1, -m \right] \), \(\tau_{10} =\left[0, -m-1, -m-1 \right] \), \(\tau_{11} =\left[1-m-1, -m, -m \right] \) and \(\tau_{12} =\left[0, -m, -m \right] \). Similarly, the effective CDF on the FSO link can be obtained by substituting (\ref{FSO-CDF}) and (\ref{Intf-R}) in \(F_{\gamma_{\scaleto{{T_2,FSO}}{4 pt}}}(\gamma) = \int_{0}^{\infty} F_{\gamma_{\scaleto{{FSO}}{4 pt}}}(\gamma y) f_{_{{\scaleto{I_{r}}{5 pt}}}}(y) dy \), which can be evaluated using \cite[Eq. (07.34.21.0013.01)]{Wolfram}. The resulting expression can be further differentiated using \cite[Eq. (07.34.20.0017.02)]{Wolfram} to express the PDF of \(\gamma_{\scaleto{{T_2,FSO}}{4 pt}}\). Plugging the PDF, thus obtained, into the relationship \({\mathcal{R}}_2 = \frac{1}{2} \int_{0}^{\infty} \text{log}_2(1+\gamma) f_{\gamma_{\scaleto{{T_2,FSO}}{4 pt}}}(\gamma) d\gamma\), the closed-form expression of \({\mathcal{R}}_2\) can be obtained as given below
	\begin{align*}
\setcounter{equation}{19}
{\mathcal{R}}_2 = &  \frac{\text{log}_2(\text{e})\mathcal{D}_4(y)^{Nm_1-\frac{3}{2}}}{2\delta\Gamma(m_1N)(2\pi)^{\frac{3}{2}(y-1)}}  \meijerG{\text{n}+2\text{y}}{3\text{y}+1}{4\text{y}+\text{r}+1}{\text{n}+3\text{y}+1}{\tau_{13}}{\tau_{14}}{ \frac{\mathcal{D}_{7}}{\delta^y} }  \label{R2-Cap} \addtag
\end{align*}
where \(\tau_{13}\) \(=\) \([0,\Delta(y:1),1,\tau_{5},\tau_{3}]\) and \(\tau_{14}\) \(=\) \([\tau_{5},\Delta(y:1),1,0]\). Finally, on substituting (\ref{R1-Cap}) and (\ref{R2-Cap}) into (\ref{ASR_def}), the expression of ASR can be established.  
Furthermore, the asymptotic approximation of Meijer-G function involved in \({\mathcal{R}}_1\) can be developed considering the fact that as \(\bar\gamma_{k} \to \infty \), the argument \(\mathcal{B}_{3} \to 0 \). Invoking \cite[07.34.06.0006.01]{Wolfram}, the Meijer-G in \({\mathcal{R}}_1\) can be approximated as:
\begin{align*}
 \meijerG{3}{2}{3}{3}{\tau_{9}}{\tau_{10}}{ \mathcal{B}_{3} } \underrel{\bar\gamma_{k} \to \infty}{\simeq} &  \sum_{i=1}^{3} \frac{\prod_{j=1, j\neq i}^{3}{\Gamma( \tau_{10,j} - \tau_{10,i} )}}{\Gamma(\tau_{9,3} - \tau_{10,i} )}{ \mathcal{B}_{3}}^{\tau_{10,i}} \\ & \times 
\prod_{j=1}^{2}{\Gamma(1- \tau_{9,j} + \tau_{10,i} )} \addtag \label{M1-Asym}
\end{align*}
Similar approach can be opted to represent \(\meijerG{3}{2}{3}{3}{\tau_{11}}{\tau_{12}}{ \mathcal{B}_{3} } \) asymptotically. Additionally, in order to simplify the expression of \({\mathcal{R}}_2\), it can be noted that as \(\mu_{\scaleto{r}{3 pt}} \to \infty \), the argument of Meijer-G in (\ref{R2-Cap}) diminishes, i.e., \(\mathcal{D}_7 \to 0 \), which can be approximated using \cite[07.34.06.0006.01]{Wolfram} as:
\begin{align*}
& \meijerG{\text{n}+2\text{y}}{3\text{y}+1}{4\text{y}+\text{r}+1}{\text{n}+3\text{y}+1}{\tau_{13}}{\tau_{14}}{ \frac{\mathcal{D}_{7}}{\delta^y} } \underrel{ \mu_{\scaleto{r}{3 pt}} \to \infty}{\simeq}   \sum_{i=1}^{\text{n}+2\text{y} } \frac{\prod_{j=1, j\neq i}^{\text{n}+2\text{y}}{\Gamma( \tau_{14,j} - \tau_{14,i} )}}{ \prod_{j=3\text{y}+2}^{4\text{y}+r+1} \Gamma(\tau_{13,j} - \tau_{14,i} )  }  \\ & \times { \left(\frac{\mathcal{D}_{7}}{\delta^y}\right) }^{\tau_{14,i}}  
\frac{\prod_{j=1}^{3\text{y}+1} {\Gamma(1- \tau_{13,j} + \tau_{14,i} )}}{ \prod_{j=\text{n}+2\text{y}+1, j\neq i}^{\text{n}+3\text{y}+1}{\Gamma( 1- \tau_{14,j} + \tau_{14,i} )} } \addtag \label{M3-Asym}
\end{align*}
Substituting the aforementioned approximations of Meijer-G function in (\ref{R1-Cap}) and (\ref{R2-Cap}), the asymptotic representation of ASR can be obtained.
\section{NUMERICAL RESULTS}
In this section, numerical examples are shown to demonstrate the findings of the research work,  together with Monte-Carlo simulations. Fig. 1 illustrates the OP of mixed RF/FSO fixed-gain AF TWR systems versus the average SNR per hop in moderate (i.e., \(\alpha_1=2.1, \alpha_2=2 , \beta_1=4 , \beta_2=4.5 ,  \Omega_1=1.0676 ,  \Omega_2=1.06 \)) turbulence conditions. The figure also investigates the effect of strong (i.e., w/a=5) and weak (i.e., w/a=10) pointing errors on the system performance. OP deteriorates by increasing the number of interferers, i.e., \(N\). It can be noted that, at high SNR, the asymptotic expansion matches very well with its exact counterpart, which confirms the validity of our mathematical analysis for different parameter settings. On the other hand, it can be observed that coherent detection  (\(r=1\)) outperforms IM/DD (\(r=2\)) in turbulent environments as previously observed. Additionally, the impact of type demodulation schemes, along with the number of RF users, on the ASR has been demonstrated in Fig. 2. It can be noted that increasing \(K\) provides a remarkable improvement in the system performance. Finally, the deterioration introduced by interference on ASR of the considered TWR relay network is quantified in Fig. 3. In the plot, the parameters assumed to model turbulence on the FSO link is given as: \(\alpha_1=2.169, \alpha_2=1 ,  \beta_1=0.55 , \beta_2=2.35 ,  \Omega_1=1.5793 ,  \Omega_2=1 \). It is evident in the plot that the average rate improves as the number of interferers, \(N\), reduces.
\section{CONCLUSION}
In this work, performance analysis for a mixed MUD-RF/FSO TWR network in the presence of interference is presented. Specifically, the exact and asymptotic closed-form expressions for OP and ASR have been derived in order to demonstrate the effect of various model parameters on overall performance of considered system. The analysis is extended to provide asymptotic OP expression. The derived results account for both IM/DD and coherent demodulation techniques on the optical link. The improvement brought about by MUD on RF link and two-way transmission is evident in the findings.

\bibliographystyle{IEEEtran}
{\small
	\bibliography{TWR_DobleG}}

\begin{thebibliography}{10}
\providecommand{\url}[1]{#1}
\csname url@samestyle\endcsname
\providecommand{\newblock}{\relax}
\providecommand{\bibinfo}[2]{#2}
\providecommand{\BIBentrySTDinterwordspacing}{\spaceskip=0pt\relax}
\providecommand{\BIBentryALTinterwordstretchfactor}{4}
\providecommand{\BIBentryALTinterwordspacing}{\spaceskip=\fontdimen2\font plus
\BIBentryALTinterwordstretchfactor\fontdimen3\font minus
  \fontdimen4\font\relax}
\providecommand{\BIBforeignlanguage}[2]{{%
\expandafter\ifx\csname l@#1\endcsname\relax
\typeout{** WARNING: IEEEtran.bst: No hyphenation pattern has been}%
\typeout{** loaded for the language `#1'. Using the pattern for}%
\typeout{** the default language instead.}%
\else
\language=\csname l@#1\endcsname
\fi
#2}}
\providecommand{\BIBdecl}{\relax}
\BIBdecl

\bibitem{E.Lee}
E.~Lee, J.~Park, D.~Han, and G.~Yoon, ``Performance analysis of the asymmetric
  dual-hop relay transmission with mixed {RF/FSO} links,'' \emph{IEEE
  Photon.Techn. Lett.}, vol.~23, no.~21, pp. 1642--1644, Nov. 2011.

\bibitem{DGGwithPointError}
H.~AlQuwaiee, I.~S. Ansari, and M.~S. Alouini, ``On the performance of
  free-space optical communication systems over {D}ouble {G}eneralized {G}amma
  channel,'' \emph{IEEE J. on Selec. Areas Commun.}, vol.~33, no.~9, pp.
  1829--1840, Sept. 2015.

\bibitem{TWR-RF}
B.~Xia, C.~Li, and Q.~Jiang, ``Outage performance analysis of multi-user
  selection for two-way full-duplex relay systems,'' \emph{IEEE Commun. Lett.},
  vol.~21, no.~4, pp. 933--936, Apr. 2017.

\bibitem{RF-FSO-Rel-Det_Intf}
E.~Soleimani-Nasab and M.~Uysal, ``Generalized performance analysis of mixed
  {RF/FSO} cooperative systems,'' \emph{IEEE Trans. Wirel. Commun.}, vol.~15,
  no.~1, pp. 714--727, Jan. 2016.

\bibitem{RF-FSO-CSI-AF}
E.~Balti and M.~Guizani, ``Mixed {RF/FSO} cooperative relaying systems with
  co-channel interference,'' \emph{IEEE Trans. Commun.}, vol.~66, no.~9, pp.
  4014--4027, Sept. 2018.

\bibitem{mypaper2}
A.~Updhya, V.~Dwivedi, and G.~Singh, ``Multiuser diversity for mixed {RF/FSO}
  cooperative relaying in the presence of interference,'' \emph{Optics
  Communications}, Feb. 2019.

\bibitem{TWR-MUD}
Y.~F. Al-Eryani, A.~M. Salhab, S.~A. Zummo, and M.~S. Alouini, ``Two-way
  multiuser mixed {RF/FSO} relaying: performance analysis and power
  allocation,'' \emph{IEEE/OSA J. of Optical Commun. and Network.}, vol.~10,
  no.~4, pp. 396--408, Apr. 2018.

\bibitem{wireleDigAloiuni}
M.~K. Simon and M.~S. Alouini, ``Digital communication over fading channels,''
  \emph{2nd ed., Wiley}, 2005.

\bibitem{RyzhikTables}
I.~S. Gradshteyn and I.~M. Ryzhik, ``Table of integrals, series and products,''
  \emph{7th ed., A. Jeffrey, Ed. Elsevier Inc.}, 2007.

\bibitem{FoxHIntgrl}
P.~Mittal and K.~Gupta, ``An integral involving generalized function of two
  variables,'' \emph{Proc. Ind. Acad. Sci.}, vol.~75, no.~3, pp. 117--123, Nov.
  1972.

\bibitem{BivFoxMATLAB}
H.~Chergui, M.~Benjillali, and M.~S. Alouini, ``Rician {K}-factor-based
  analysis of {XLOS} service probability in 5{G} outdoor ultra-dense
  networks,'' \emph{IEEE Wirel. Commun. Lett.}, 2018.

\bibitem{ResTheom}
A.~Kilbas, ``{H}-transforms: Theory and applications. analytical methods and
  special functions,'' \emph{Taylor and Francis}, 2004.

\bibitem{Wolfram}
I.~Wolfram, ``Wolfram, research, mathematica edition: Version 10.0.
  champaign,'' \emph{Wolfram Research, Inc.}, 2010.

\end{thebibliography}
%\bibliography{alpha_ref4}

\end{document}